\documentclass[sigconf]{acmart}
\AtBeginDocument{%
  }

\usepackage{algorithmic}
\usepackage{graphicx}
\usepackage{textcomp}
\usepackage{xcolor}
\usepackage{balance}
\usepackage{xspace}
\usepackage{pifont}
\usepackage{subcaption}
\usepackage{booktabs}
\usepackage{colortbl}
\usepackage{multirow}
\usepackage{url}
\usepackage{array,ragged2e}
\usepackage{booktabs}
\usepackage{enumitem}
\usepackage{tablefootnote}

\setcopyright{none} 
\renewcommand{\footnotetextcopyrightpermission}[1]{} 

\newcommand{\ie}{\emph{i.e.,}\xspace}
\newcommand{\eg}{\emph{e.g.,}\xspace}

\newcommand{\etal}{\emph{et~al.}\xspace} 

\begin{document}

\title{Quality Model for Machine Learning Components}

\author{Grace A. Lewis}
\email{glewis@sei.cmu.edu}
\orcid{0000-0001-9128-9863}
\affiliation{%
  \institution{Carnegie Mellon Software Engineering Institute}
  \city{Pittsburgh}
  \state{PA}
  \country{USA}
}

\author{Rachel Brower-Sinning}
\email{rbrowersinning@sei.cmu.edu}
\orcid{0009-0008-9280-7675}
\affiliation{%
  \institution{Carnegie Mellon Software Engineering Institute}
  \city{Pittsburgh}
  \state{PA}
  \country{USA}
}

\author{Robert Edman}
\email{rbedman@sei.cmu.edu}
\orcid{0000-0002-1293-2417}
\affiliation{%
  \institution{Carnegie Mellon Software Engineering Institute}
  \city{Pittsburgh}
  \state{PA}
  \country{USA}
}

\author{Ipek Ozkaya}
\email{ozkaya@sei.cmu.edu}
\orcid{0000-0002-7336-4775}
\affiliation{%
  \institution{Carnegie Mellon Software Engineering Institute}
  \city{Pittsburgh}
  \state{PA}
  \country{USA}
}

\author{Sebastián Echeverría}
\email{secheverria@sei.cmu.edu}
\orcid{0000-0003-1144-8612}
\affiliation{%
  \institution{Carnegie Mellon Software Engineering Institute}
  \city{Pittsburgh}
  \state{PA}
  \country{USA}
}

\author{Alex Derr}
\email{aderr@sei.cmu.edu}
\orcid{0009-0003-3580-210X}
\affiliation{%
  \institution{Carnegie Mellon Software Engineering Institute}
  \city{Pittsburgh}
  \state{PA}
  \country{USA}
}

\author{Collin Beaudoin}
\email{cbeaudoin@fairfield.edu}
\orcid{0009-0005-7766-1223}
\affiliation{%
  \institution{Fairfield University}
  \city{Fairfield}
  \state{CT}
  \country{USA}
}

\author{Katherine R. Maffey}
\email{kmaffey@andrew.cmu.edu}
\orcid{0000-0002-1677-6343}
\affiliation{%
  \institution{Carnegie Mellon University}
  \city{Pittsburgh}
  \state{PA}
  \country{USA}
}

\renewcommand{\shortauthors}{Lewis et al.}

\begin{abstract}
Despite increased adoption and advances in machine learning (ML), there are studies showing that many ML prototypes do not reach the production stage and that testing is still largely limited to testing model properties, such as model performance, without considering requirements derived from the system it will be a part of, such as throughput, resource consumption, or robustness. This limited view of testing leads to failures in model integration, deployment, and operations. In traditional software development, quality models such as ISO 25010 provide a widely used structured framework to assess software quality, define quality requirements, and provide a common language for communication with stakeholders. A newer standard, ISO 25059, defines a more specific quality model for AI systems. However, a problem with this standard is that it combines system attributes with ML component attributes, which is not helpful for a model developer, as many system attributes cannot be assessed at the component level. In this paper, we present a quality model for ML components that serves as a guide for requirements elicitation and negotiation and provides a common vocabulary for ML component developers and system stakeholders to agree on and define system-derived requirements and focus their testing efforts accordingly. The quality model was validated through a survey in which the participants agreed with its relevance and value. The quality model has been successfully integrated into an open-source tool for ML component testing and evaluation demonstrating its practical application.
\end{abstract}




\maketitle

\section{Introduction}\label{sec:introduction}

It is well known that the adoption of machine learning (ML) is increasing and with it the availability of ML model development tools and MLOps infrastructures to ease and close the gap between ML model development, deployment, and operations \cite{Ozkan2024}\cite{Schwaeke2025}. However, despite advances in automation, only \textasciitilde40\% of ML prototypes reach production according to Gartner \cite{Gartner2025}, and testing is largely limited to testing model properties, such as model performance (\eg accuracy) \cite{Hutchinson2022}, without considering requirements derived from the system that it will be a part of, such as throughput, resource consumption, or robustness. This limited view of testing leads to failures in model integration, deployment, and operations \cite{Chandrasekaran2024}\cite{Stone2025}. Although in some cases model developers may not have the right skills to test beyond accuracy, in many cases the system context and requirements are not shared with model developers, as ML components are often not treated as software components. In addition, in many organizations, model developers do not belong to the product team, or model development is outsourced, further exacerbating collaboration challenges that result in poor or uninformed testing practices \cite{Lewis2021a}\cite{Lewis2021b}\cite{Nahar2023}. 

In traditional software development, quality models such as ISO 25010:2023 provide a widely-used structured framework to assess software quality, define quality requirements, and provide a common language for communication with stakeholders \cite{ISO25010-2023}. ISO 25059:2023 defines a quality model for AI systems based on the 2010 version of the ISO 25010 standard \cite{ISO25059-2023}. However, an observed problem with this standard is that it combines system attributes with ML component attributes, which is not helpful for a model developer, as many cannot be assessed at the component level \cite{Ali2022}. For example, \textit{co-existence} is a quality attribute defined by ISO 25010 and ISO 25059 as "Capability of a product to perform its required functions efficiently while sharing a common environment and resources with other products, without detrimental impact on any other product." It is not possible for an ML component developer to evaluate this quality attribute independently; it can only be assessed at the system level. However, co-existence as a system-level quality attribute can impose requirements on the ML component such as time behavior or resource utilization. Test and evaluation at the ML component level is especially important in situations in which model development occurs in a separate team (\eg data science team) or is contracted to an outside organization, or a pre-trained model is being assessed for internal use, which are all very common use cases and scenarios in industry \cite{Banyongrakkul2025}\cite{Nahar2022}.

The main contribution of this paper is a quality model for ML components that serves as a guide for requirements elicitation and negotiation and provides a common vocabulary for ML component developers and system stakeholders to agree on and define system-derived requirements and focus their testing efforts accordingly. The relevance of the quality model was validated through an invitation-only practitioner survey in which participants agreed that consideration of quality attributes in the test and evaluation of ML components leads to early detection of a larger and more diverse set of problems that would otherwise not be detected until production. The quality model has been integrated into MLTE (ML Test and Evaluation)\footnote{\url{https://github.com/mlte-team/mlte}}, an open-source process and tool for ML component testing and evaluation.  The quality model guides the model developer in the elicitation of system-derived requirements. In addition, MLTE includes a test catalog that serves as an organizational repository of test code examples, which is organized using the same quality model, such that model developers can easily locate reusable test code. The default test catalog included with MLTE includes at least one example for each quality attribute in the model.

The paper is organized as follows. Section \ref{sec:background} introduces terminology that sets the context for the paper. The methodology used to develop the quality model is presented in Section \ref{sec:methodology}, along with the proposed quality model. The results of the survey to validate the relevance of the quality model in practice are presented in Section \ref{sec:evaluation}. Section \ref{sec:discussion} presents a series of observations based on the findings of the survey. Finally, Section \ref{sec:threats} presents threats to validity of our findings, Section  \ref{sec:related-work} describes related work, and Section \ref{sec:summary} summarizes the paper and outlines next steps for our work.

\section{Terminology}\label{sec:background}

In this section, we define terminology that serves as context for the proposed quality model as well as an attempt to start creating a shared vocabulary for developers of ML components and ML-enabled systems as well as researchers and others working on practices and standards to drive ML-related  quality efforts.

\subsection{ML-Enabled System}

\begin{figure*}[h]
  \centering
  \includegraphics[width=15cm]{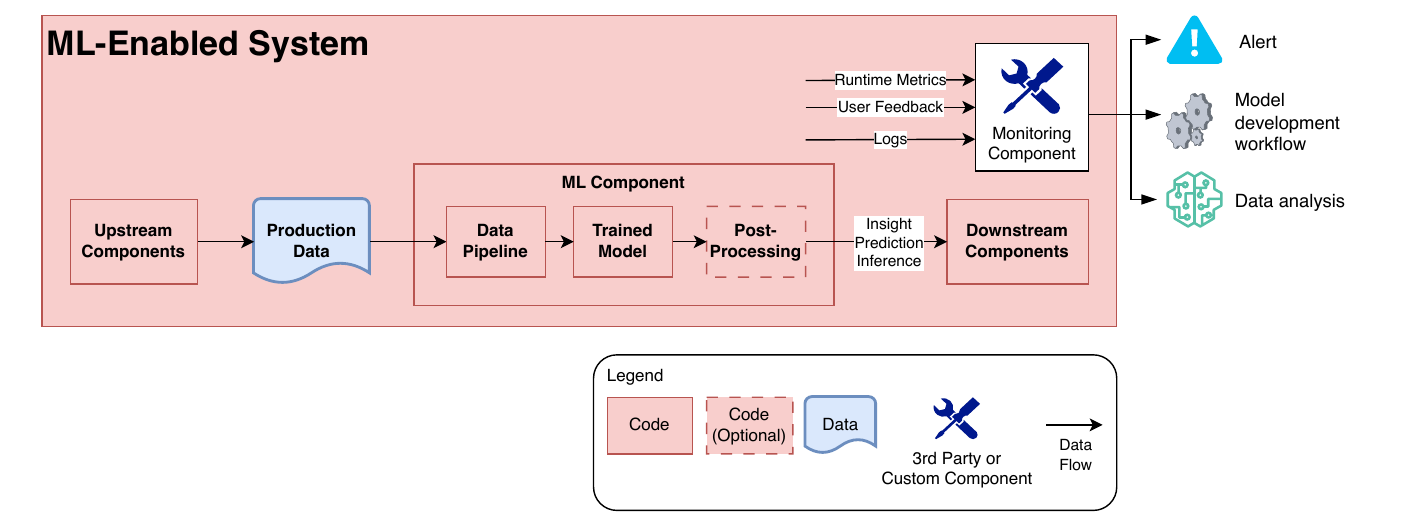}
  \caption{ML-Enabled System}
  \label{fig:ml-enabled-system}
\end{figure*}

As shown in Figure \ref{fig:ml-enabled-system}, we define an \textbf{ML-Enabled System} as a software system that relies on one or more ML-based components to provide required capabilities. Within the system, we explicitly recognize the need for a \textbf{Monitoring Component} to account for potential non-deterministic behavior and produce alerts or triggers when the component reaches pre-defined thresholds associated to monitored metrics and artifacts (\eg drift conditions, boundary conditions, user feedback, logs).

\subsection{ML Component} 

As shown in Figure \ref{fig:ml-enabled-system}, we define an \textbf{ML Component} as the \textbf{Data Pipeline} that transforms \textbf{Production Data} from \textbf{Upstream Components} into model inputs and the \textbf{Trained Model} (or Pre-Trained Model) that produces an output that is consumed by \textbf{Downstream Components}. Optionally, there may be a \textbf{Post-Processing} sub-component to translate model outputs to conform to APIs or data formats required by Downstream Components.

\begin{figure}[h]
  \centering
  \includegraphics[width=\linewidth]{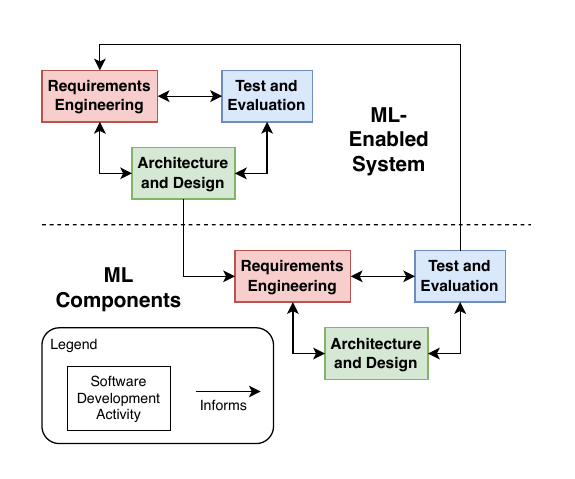}
  \caption{ML Component Requirement Sources (adapted from Kuwajima \etal \cite{Kuwajima2020})}
  \label{fig:ml-component-requirements}
\end{figure}

As shown in Figure \ref{fig:ml-component-requirements} (adapted from Kuwajima \etal \cite{Kuwajima2020}), system decomposition into components occurs during system-level architecture and design. Some components will be identified as ML components and derive requirements from the system. Even for pre-trained models, or in cases where the system is built around a model (\ie model-first development \cite{Nahar2023}), the ML component will have \textit{system-derived requirements}. The results of testing the ML component against these requirements will inform system-level activities, especially in the case of unrealistic requirements placed on ML components.

\subsection{Quality Model}

We use the term quality model in the same manner as ISO 25010 \cite{ISO25010-2023}, which defines software product quality through a set of characteristics that provide a consistent way to specify, measure, and evaluate its quality. In this paper, the developed quality model defines ML component quality through a set of characteristics that correspond to \textit{quality attributes} (QAs) as defined by Bass \etal \cite{Bass2022} and adapted to ML components: a measurable or testable property of a component that is used to indicate how well the component satisfies its system-derived requirements beyond the basic function of the component.

\section{Methodology}\label{sec:methodology}

\subsection{Starting Point}
\label{subsec:starting-point}

As the goal was not to recreate existing work but to fill the gap in quality attributes of ML components and their relationship with system-derived requirements, our starting point was the ISO 25010:2023 and ISO 25029:2023 quality models for software products and AI systems, respectively \cite{ISO25010-2023}\cite{ISO25059-2023}. Because ISO 25029 covers both product software quality and quality-in-use, we added the ISO 
25019:2023 quality-in-use model \cite{ISO25019-2023} so that we could have both the source quality models (25010 and 25019) plus the derived quality model for AI systems (25059).

To complement the standards with the perspective of similar work in the existing literature, a team of three authors (two data scientists and one software engineer) conducted a literature search to find studies that proposed quality models for ML systems. We searched in Google Scholar for variations (\eg singular/plural, acronym) of the keywords machine learning, quality model, quality attribute, and non-functional requirement, and identified a total of 19 papers that were in English and we had access to the full text. Each paper was fully read by the three authors and individually evaluated against inclusion and exclusion criteria to ensure that it proposed a quality model that considered a comprehensive set of relevant QAs for ML-enabled systems and/or ML components (similar to the standards documents). Any disagreements were resolved by discussing the paper against the inclusion criteria. This process identified three relevant studies (as of November 2024) and is documented in the replication package \footnote{\url{https://doi.org/10.5281/zenodo.17419045}}:

\begin{itemize}
    \item \textit{Best Practices for Machine Learning Systems: An Industrial Framework for Analysis and Optimization} by Chouliaras \etal \cite{Chouliaras2023}
    \item \textit{Non-Functional Requirements For Machine Learning: Understanding Current Use and Challenges Among Practitioners} by Habibullah \etal \cite{Habibullah2023}
   \item \textit{Requirements-Driven Method to Determine Quality Characteristics and Measurements for Machine Learning Software and its Evaluation} by Nakamichi \etal \cite{Nakamichi2020}
\end{itemize}

Unfortunately, we had to eliminate the third study because the list of QAs was only a partial list.  We reached out to the authors and regrettably did not receive a response. However, the first study represents a position from industry (booking.com), and the second a position from academia, which provides two perspectives.

\subsection{Quality Model Development Process}
\label{subsec:process}

\begin{figure*}[h]
  \centering
  \includegraphics[width=\linewidth]{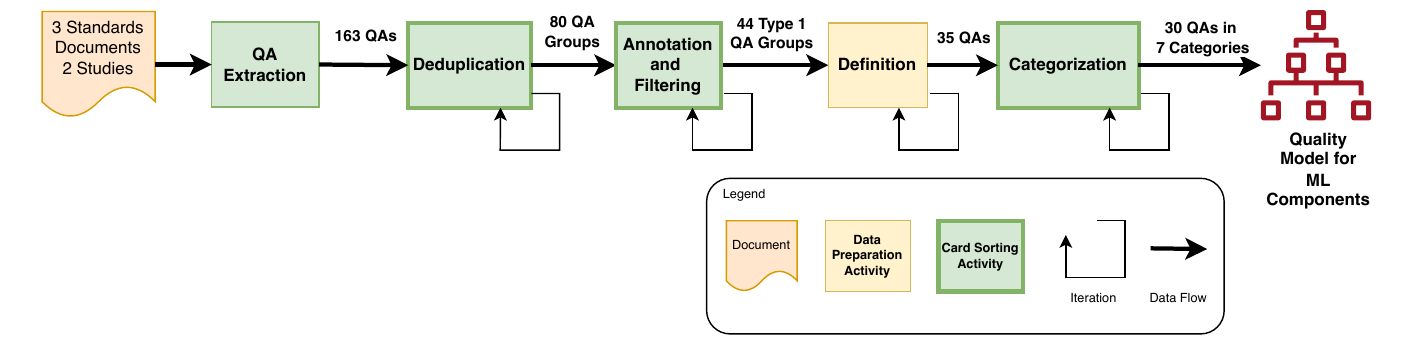}
  \caption{Quality Model Development Process}
  \label{fig:card-sorting}
\end{figure*}

The process by which the quality model was developed consists of five activities and is shown in Figure~\ref{fig:card-sorting}. All data and code produced and used during the process are available in the replication package. Four team members (two with a software engineering/software architecture background and two with a data science/ML background) executed the process, with a team member joining at the end for the final categorization.

In the \textit{QA Extraction} activity, 163 QAs were extracted verbatim from the three standards documents and the two secondary studies listed in Section~\ref{subsec:starting-point}.  We used card sorting, a common technique for grouping information, to derive the QAs and categories of the quality model \cite{Hudson2012}\cite{Spencer2009}. Following the standard card sorting method, we created one card per QA and incrementally and iteratively organized those cards into groups of similar QAs. Each card contained the definition of the QA, source document, and the category assigned by the document.

In the \textit{Deduplication} activity the goal was for participants to become familiar with the QAs and eliminate duplicates given the large amount of cards, by placing duplicate QAs in the same group. Not surprisingly, in addition to duplication between sources, the team found that different sources used different names for the same QA. After several iterations, we ended up with 80 QA groups, with the smallest groups containing one card and the largest group containing seven cards. 

In the \textit{Annotation and Filtering} activity, each participant annotated each QA group with a type:

\begin{itemize}
    \item \textbf{Type 1.} QA applies to the ML component, which means that it can be tested at the component level.
    \item \textbf{Type 2.} QA applies to the ML-enabled system, but places requirements on the ML component (\ie system-derived requirement).
    \item \textbf{Type 3.} QA applies strictly to the ML-enabled system, which means that it can only be tested at the system level.
\end{itemize}

 Each group with at least one disagreement was discussed. Each participant explained their annotation, and as a team we analyzed the rationale against the type definitions. After several iterations and discussions, 44 QA groups (55\%) were annotated as Type 1, 65 QA groups as Type 2 (82\%), and 10 QA groups as Type 3 (13\%). It is important to note that a QA group could be annotated as Type 1 and Type 2 (\eg monitoring, fairness, time behavior); our groups had 4 QA groups marked exclusively as Type 1 and 40 groups marked as Type 1 and Type 2. We initially achieved moderate inter-rater agreement with Fleiss-kappa = 0.539 and Light-kappa = 0.541 \cite{Han2024}. However, this being a collaborative effort between team members of different backgrounds, we did not expect a high inter-rater agreement and instead worked together as a team to build consensus.

 In the \textit{Definition} activity, for the 44 QA groups marked as Type 1 (including those marked Type 1 and 2), a name and definition focused on ML capabilities were produced by one of the participants based on the information in the cards for that group. This involved extracting key aspects of each definition in the QA group and placing them in the context of ML components. Further discussion was conducted asynchronously, with one participant consolidating the information and feedback and sending it back for further review. We conducted three asynchronous rounds until there was agreement on 35 distinct QAs and their definitions that were ready for categorization.

 In the final \textit{Categorization} activity, we came together again as a team to create categories for the identified ML component QAs using standard card sorting. We created new cards with the 35 identified QA names and definitions. The exercise resulted in 9 categories and 34 quality attributes (\textit{Trustworthiness} was removed after discussions concluded that it was a composite property based on multiple QAs). Further discussion was conducted asynchronously, with one participant consolidating the information and feedback and sending it back for further review. We added a fifth team member to bring a new perspective to the discussion and help with disagreements. We conducted three asynchronous rounds until there was minimal disagreement between the participants, at which point we scheduled a final in-person meeting to resolve the disagreements and visualize the categorization as a whole. The result is a quality model with 30 QAs grouped into 7 categories that represent testable properties of ML components.

\begin{figure*}[h]
  \centering
  \includegraphics[width=16cm]{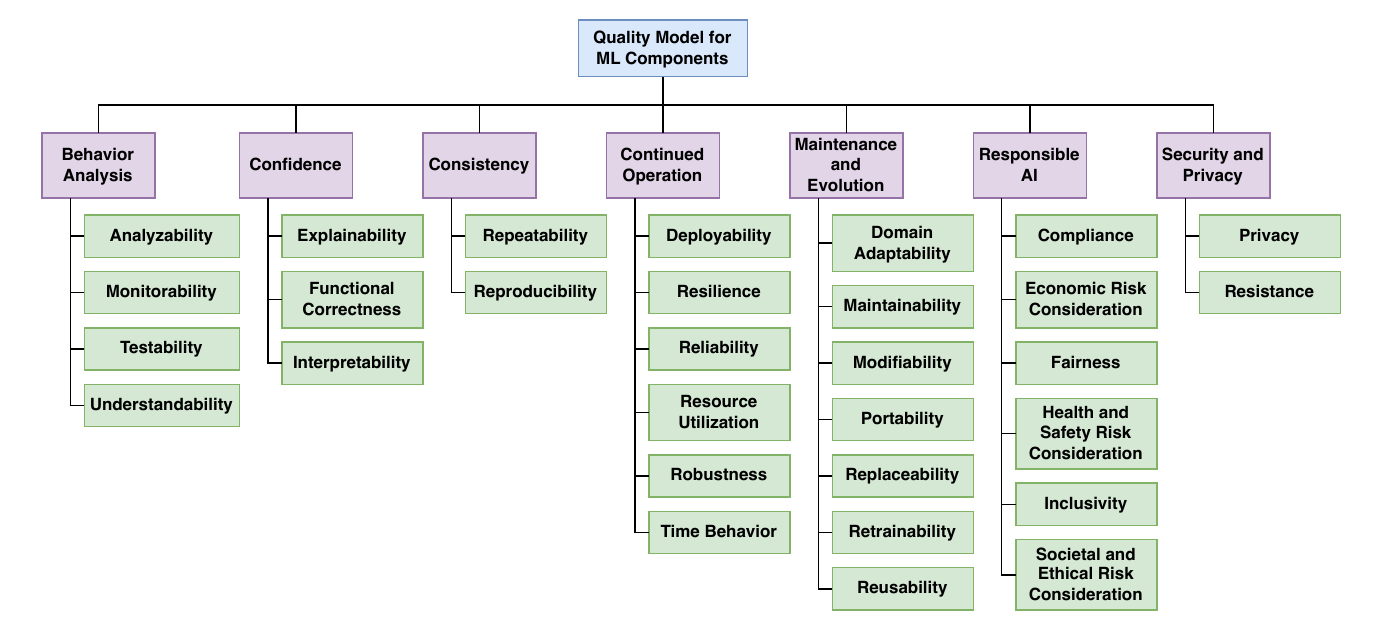}
  \caption{Quality Model for ML Components}
  \label{fig:quality-model}
\end{figure*}

 The quality model is shown in Figure \ref{fig:quality-model}. The definitions for each category and QA are provided in Table \ref{tab:quality-model-defs-1} and Table \ref{tab:quality-model-defs-2}.

\begin{table*}[t]
  \caption{Quality Model Definitions --- Part 1}
  \label{tab:quality-model-defs-1}
  \begin{tabular}{p{2.3cm}p{14.8cm}}
    \toprule
    \multicolumn{2}{c}{\textbf{Behavior Analysis:} QAs related to ease of observing and analyzing ML component behavior at development time and run time.}\\
    \midrule
    Analyzability & Capability of an ML component to expose attributes that enable it to be effectively and efficiently assessed regarding its behavior or the impact of an intended change, to diagnose it for deficiencies or causes of failure, or to identify sub-components to be modified.\\
    Monitorability & Capability of an ML component to produce data that  can be used to measure runtime attributes (\eg performance degradation, latency) that can effectively be observed and used by the ML-enabled system or operator to take action aligned with requirements.\\
    Testability & Capability of an ML component to support testing against expected behaviors by offering information relevant to test results or ensuring the visibility of failures.\\
    Understandability & Ease with which the code, implementation choices, and design choices of an ML component can be understood.\\
    \midrule
    \multicolumn{2}{c}{\textbf{Confidence:} QAs related to producing information that increases insight into the model and its outputs.}\\
    \midrule
    Explainability & Capability of an ML component to provide information that can explain model outputs (decisions) in human terms.\\
    Functional \mbox{Correctness} & Capability of an ML component to meet system-derived correctness or model performance requirements.\\
    Interpretability & Capability of an ML component to be transparent in its execution and produce information concerning the relationship between inputs and outputs.\\
    \midrule
    \multicolumn{2}{c}{\textbf{Consistency:} QAs related to producing consistent results over time.}\\
    \midrule
    Repeatability & Capability of an ML component to produce equivalent results when run on equivalent inputs during inference time.\\
    Reproducibility & Capability of the training algorithm for the ML model to produce functionally-equivalent ML models when run on similar datasets during training time.\\
    \midrule
    \multicolumn{2}{c}{\textbf{Continued Operation:} QAs related to providing continued operation within specified boundaries and constraints.}\\
    \midrule
    Deployability & Capability of an ML component to be deployed into production when needed, without any unanticipated side effects and within specified resource and time constraints.\\
    Resilience & Capability of an ML component to provide and maintain an acceptable level of service in the face of technical challenges to normal operation.\\
    Reliability & Capability of an ML component to perform specified functions without failure under normal operation.\\
    Resource \mbox{Utilization} & Capability of an ML component to use no more than the specified amount of resources to perform its function.\\
    Robustness & Capability of ML component to preserve its level of functional correctness under specified fault modes or conditions.\\
    Time \mbox{Behavior} & Capability of an ML component to produce outputs within required response time and throughput rates.\\
    \midrule
    \multicolumn{2}{c}{\textbf{Maintenance and Evolution:} QAs related to ease of maintaining and evolving ML components over time.}\\
    \midrule
    Domain \mbox{Adaptability} & Capability of an ML component trained on a source domain to be used in a different, but related domain.\\
    Maintainability & Ease with which the ML component can be modified to correct faults, to improve performance or other attributes, or to adapt to a changed environment.\\
    Modifiability & Capability of an ML component to be changed without introducing defects, unintended side effects, or degrading existing component quality, and within specified resource and time constraints.\\
    Portability & Capability of an ML component to be  adapted for or transferred to different hardware, software, or compute platform, without any unanticipated side effects and within specified resource and time constraints.\\
    Replaceability & Capability of an ML component to replace (or be replaced by) another ML component for the same purpose in the same environment.\\
    Retrainability & Ability to generate a new version of the trained model using the same process on a new set of training data, either manually or automatically depending on system-derived requirements.\\
    Reusability & Capability of an ML component (or its sub-components) to be used in other ML-enabled systems with minimal change.\\
    \bottomrule
  \end{tabular}
\end{table*}

\begin{table*}[t]
  \caption{Quality Model Definitions --- Part 2}
  \label{tab:quality-model-defs-2}
  \begin{tabular}{p{2.5cm}p{14.6cm}}
    \toprule
    \multicolumn{2}{c}{\textbf{Responsible AI:} QAs related to the ethical and societal impact of ML component usage.}\\
    \midrule
    Compliance & Capability of an ML component to satisfy requirements, as required by rules, regulations, laws, standards, or security policies.\\
    Economic Risk \mbox{Consideration} & Capability of an ML component to mitigate the potential risk of its results to financial status, efficient operation, commercial property, reputation, employment, educational status, or other resources.\\
    Fairness & Capability of an ML component to produce results that are not biased towards certain populations, including, but not limited to protected attributes (\eg race, gender, income).\\
    Health and Safety Risk Consideration & Capability of an ML component to mitigate the potential risk of its results to the health and safety of system users.\\
    Inclusivity & Capability of an ML component to be used by systems serving people of various backgrounds, characteristics, or capabilities.\\
    Societal and Ethical Risk Consideration & Capability of an ML component to mitigate the potential risk of its results to society.\\
    \midrule
    \multicolumn{2}{c}{\textbf{Security and Privacy:} QAs related to protection and handling of sensitive information.}\\
    \midrule
    Privacy & Capability of an ML component to produce outputs that cannot be mapped to personally-identified information (PII) in inputs.\\
    Resistance & Capability of an ML component to safeguard against malicious purposes.\\
    \bottomrule
  \end{tabular}
\end{table*}

\section{Evaluation}\label{sec:evaluation}

\begin{figure}[htbp]
  \centering
  \begin{subfigure}[t]{0.23\textwidth}
    \centering
    \includegraphics[width=\textwidth]{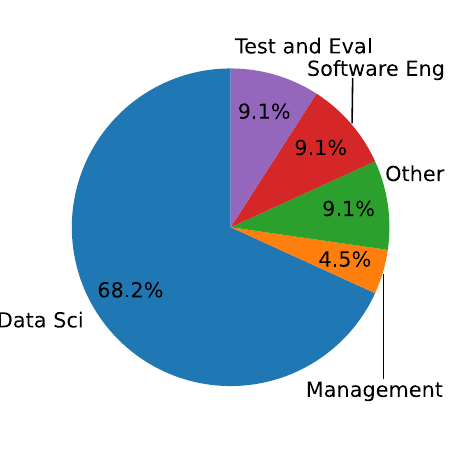}
    \caption{Organizational Role}
    \label{fig:org-role}

  \end{subfigure}
  \hfill
  \begin{subfigure}[t]{0.23\textwidth}
    \centering
    \includegraphics[width=\textwidth]{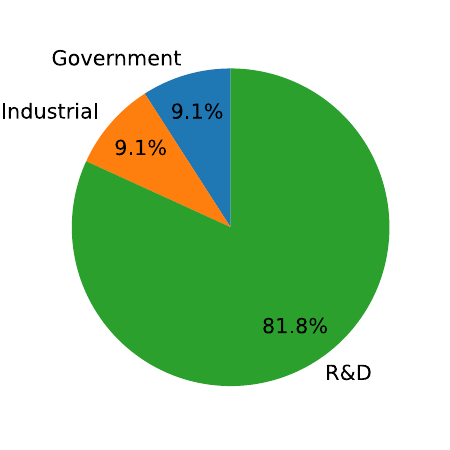}
    \caption{Organization Type}
    \label{fig:org-type}
  \end{subfigure}
  \begin{subfigure}[t]{0.23\textwidth}
    \centering
    \includegraphics[width=\textwidth]{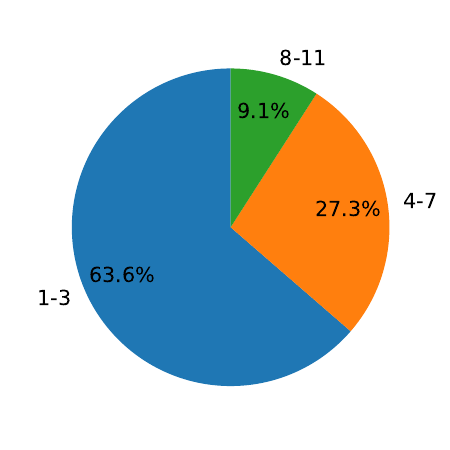}
    \caption{Years in Role}
    \label{fig:years-in-role}  
  \end{subfigure}
  \hfill
  \begin{subfigure}[t]{0.23\textwidth}
    \centering
    \includegraphics[width=\textwidth]{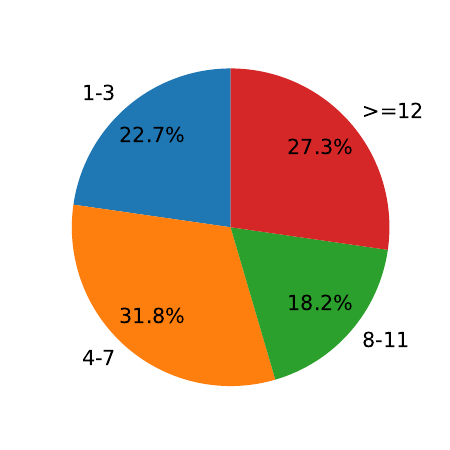}
    \caption{Total Years of Experience}
    \label{fig:years-experience}
  \end{subfigure}
  \begin{subfigure}[t]{0.23\textwidth}
    \centering
    \includegraphics[width=\textwidth]{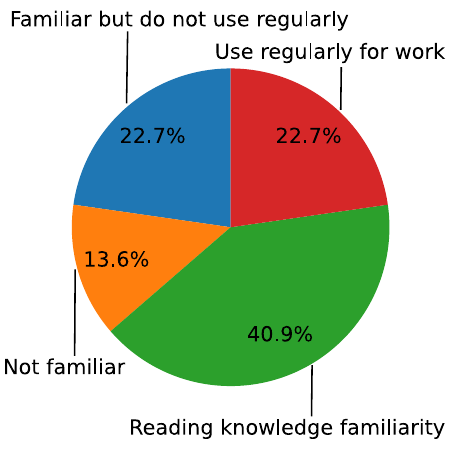}
    \caption{Familiarity with QAs}
    \label{fig:sub3}
  \end{subfigure}
  \hfill
  \begin{subfigure}[t]{0.23\textwidth}
    \centering
    \includegraphics[width=\textwidth]{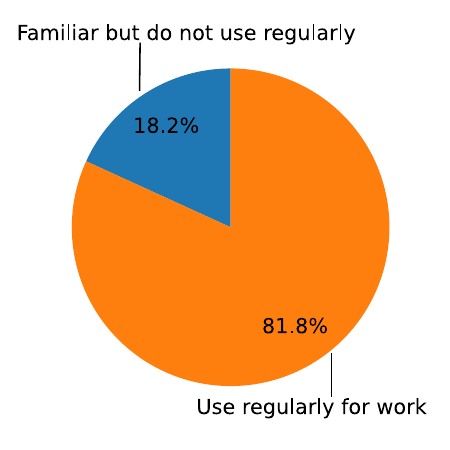}
    \caption{Familiarity with ML Component Testing}
    \label{fig:sub4}
  \end{subfigure}
  \caption{Survey Participant Demographics}
  \label{fig:demographics}
\end{figure}

An online survey was conducted to validate the quality model for ML components. The survey was conducted in August and September 2025. Participants were selected from organizations that included a government technology transition center, a software-focused research institute, and a major industrial organization. The survey received 22 complete responses. As part of the survey, demographic data from the participants was collected to provide context to the results. The complete demographic breakdown is presented in Figure~\ref{fig:demographics} and shows the diversity in organizational role, organization type, years in role, years of experience, experience with quality attributes, and familiarity with ML component testing. Notable is that Figure \ref{fig:demographics}(e) shows that 54.4\% of the participants are not familiar or have only reading knowledge of quality attributes. This supports our observation that quality attributes are not common knowledge among ML model developers, further strengthening the contribution of the quality model.  The survey instrument, anonymized responses, and analyzed data are available in the replication package. The specific research questions addressed by the survey are the following:


\begin{itemize}
    \item RQ1: What ML component QAs are currently tested in practice?
    \item RQ2: How do tested ML component QAs vary between traditional ML models and large language models (LLMs)? 
    \item RQ3: What is the importance and difficulty of testing the QAs in the proposed quality model? 
\end{itemize}


\subsection{ML Component QAs Tested in Practice (RQ1)}
\label{sec:survey-rq1}

To answer RQ1 to identify which ML component QAs are currently tested in practice, we collected open-ended responses to the question: "In your experience, what properties (\ie non-functional requirements, quality attributes) do you (or should you) usually test for ML models?"  

\begin{figure}[h]
  \centering
  \includegraphics[width=\linewidth]{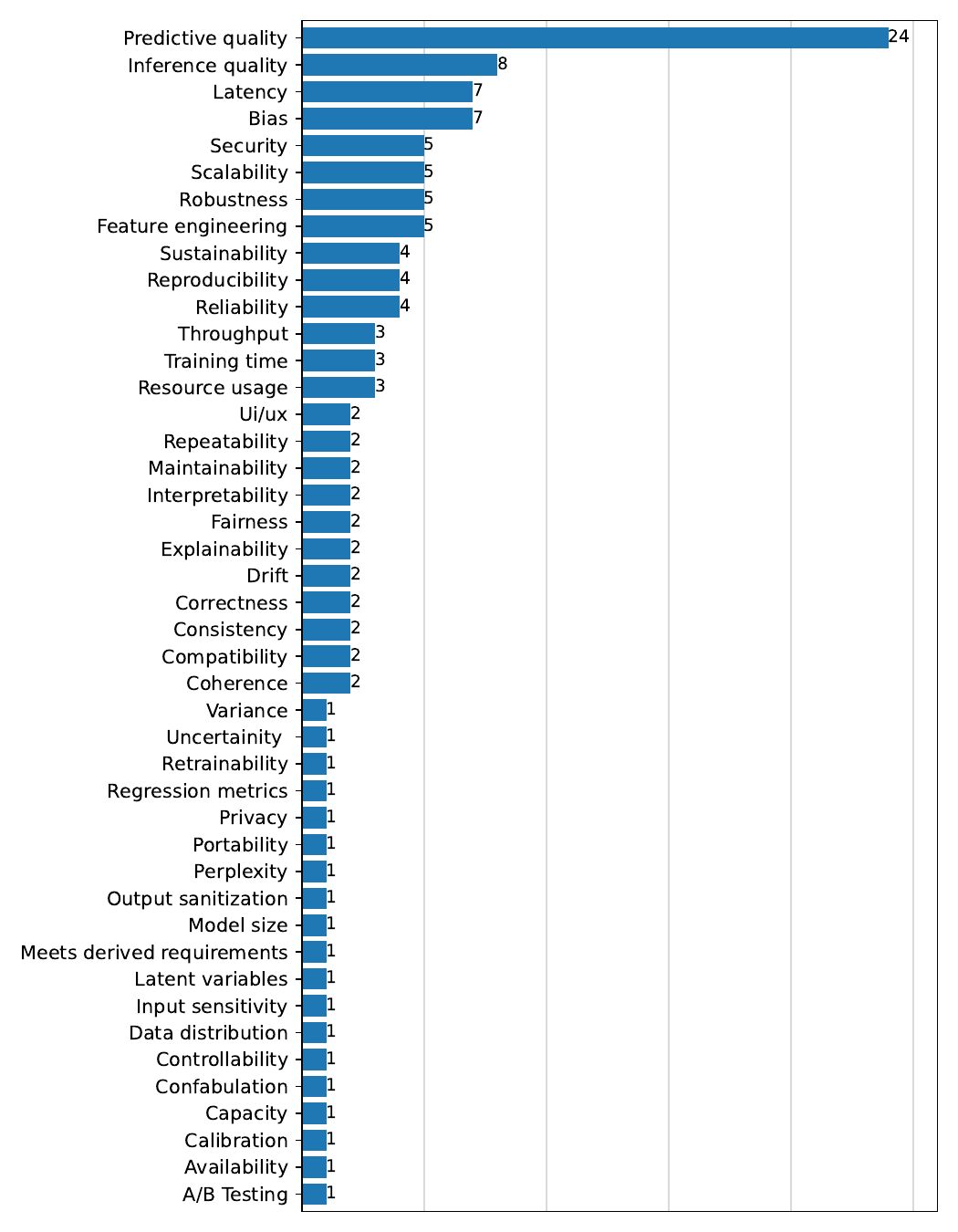}
  \caption{State of the Practice for Testing ML Components} 
  \label{fig:RQ1-additionalQAs} 
\end{figure}

We received 128 total and 70 unique answers to that question. Based on domain knowledge, one of the authors clustered the unique responses into QA groups of equivalent concepts, including those with slight language variations. For example, \textit{resource}, \textit{resource usage}, and \textit{tokens used} were clustered as \textit{resource usage}. Similarly, \textit{accuracy, weighted accuracy, f1, recall} and other accuracy related metrics were all clustered under \textit{predictive quality}. A second author reviewed the clusters for consistency. The exercise resulted in the 44 QAs shown in Figure ~\ref{fig:RQ1-additionalQAs} where the numbers on the bars correspond to the number of times the property was mentioned in the open-ended responses. As seen in this figure, most QAs were mentioned only once (\eg  availability, privacy, retrainability) or twice (\eg drift, fairness, maintainability). The QAs that were mentioned most frequently, by far, were related to predictive quality, more specifically accuracy, which constitutes \textasciitilde19\% of all responses. This was followed by inference quality, latency, and bias, each corresponding to \textasciitilde6\% of all responses. 

The overwhelming dominance of predictive quality as the most tested property validates the main motivation of this work which is that most ML testing practices today are focused on prediction quality and, therefore, highlights the important gap that our quality model fills. Although predictive quality, including accuracy, is important, not taking into account the other relevant QAs is likely to result in ML models that do not reach production successfully, as discussed in Section ~\ref{sec:introduction}. 


\subsection{Comparing QAs Tested for Traditional ML Models and LLMs (RQ2)}
\label{sec:survey-rq2)}
For RQ2, our goal was to understand whether there is a difference in the most relevant QAs tested for traditional ML models and LLMs. To address this question, we performed a further analysis of the open-ended responses, which also asked participants to specify for each property whether it applied to traditional ML models, LLMs, or both.

For the 44 QA groups listed in Figure \ref{fig:RQ1-additionalQAs}, the participants reported that most of them need to be tested for both traditional ML models and LLMs (70\%). There were a number of QA groups that the participants stated would be more relevant for traditional ML models or LLMs. Only six QA groups (14\%) were listed as more relevant to traditional ML models and included \textit{latent variables, meeting derived requirements, portability, interpretability, repeatability, and data distribution}. Similarly, there were seven attributes (16\%) listed as more relevant for LLMs, which were \textit{calibration, confabulation, controllability, output sanitization, perplexity, coherence, and compatibility}. It is possible to argue that these categorizations were a reflection of the experience of the participants and all can be relevant to both, or some attributes are less meaningful for LLMs, such as feature engineering. Regardless, the overwhelming agreement that many QAs we collected through open responses were considered applicable to testing of traditional ML models and LLMs suggests that quality models such as ours can provide broader guidance for testing ML components, including traditional ML models and LLMs.

\begin{figure}[h]
 \centering
 \includegraphics[width=\linewidth]{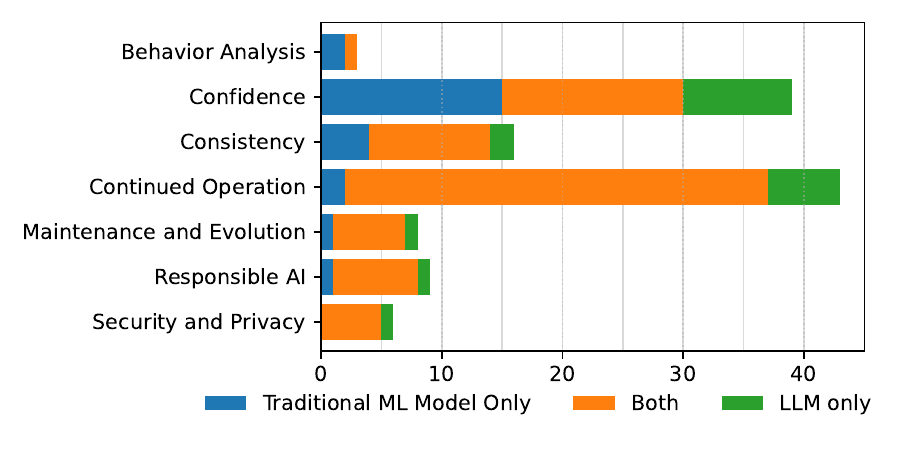}
 \caption{Testing Practices for Traditional ML Models and LLMs} 
 \label{fig:RQ1.2-testing_differences} 
\end{figure}

Additionally, to better understand the differences in testing practice, we mapped each of the 128 responses to the high level categories of the proposed quality model (Figure~\ref{fig:quality-model}). The results in Figure~\ref{fig:RQ1.2-testing_differences} show that most of the current testing practice is focused on the \textit{Confidence} and \textit{Continued Operation} categories and that most of the testing practices apply to both LLMs and Traditional ML models. Something to note from this figure is that no testing practices exclusively for LLMs were reported for \textit{Behavior Analysis} category and that no testing practices exclusively for traditional ML models were reported for the \textit{Security and Privacy} category. This analysis further confirms that the proposed quality model applies to a broad category of ML models.


%

%
\begin{figure*}[htbp]
  \centering
  \begin{subfigure}[t]{0.45\textwidth}
    \centering
    \includegraphics[width=\textwidth]{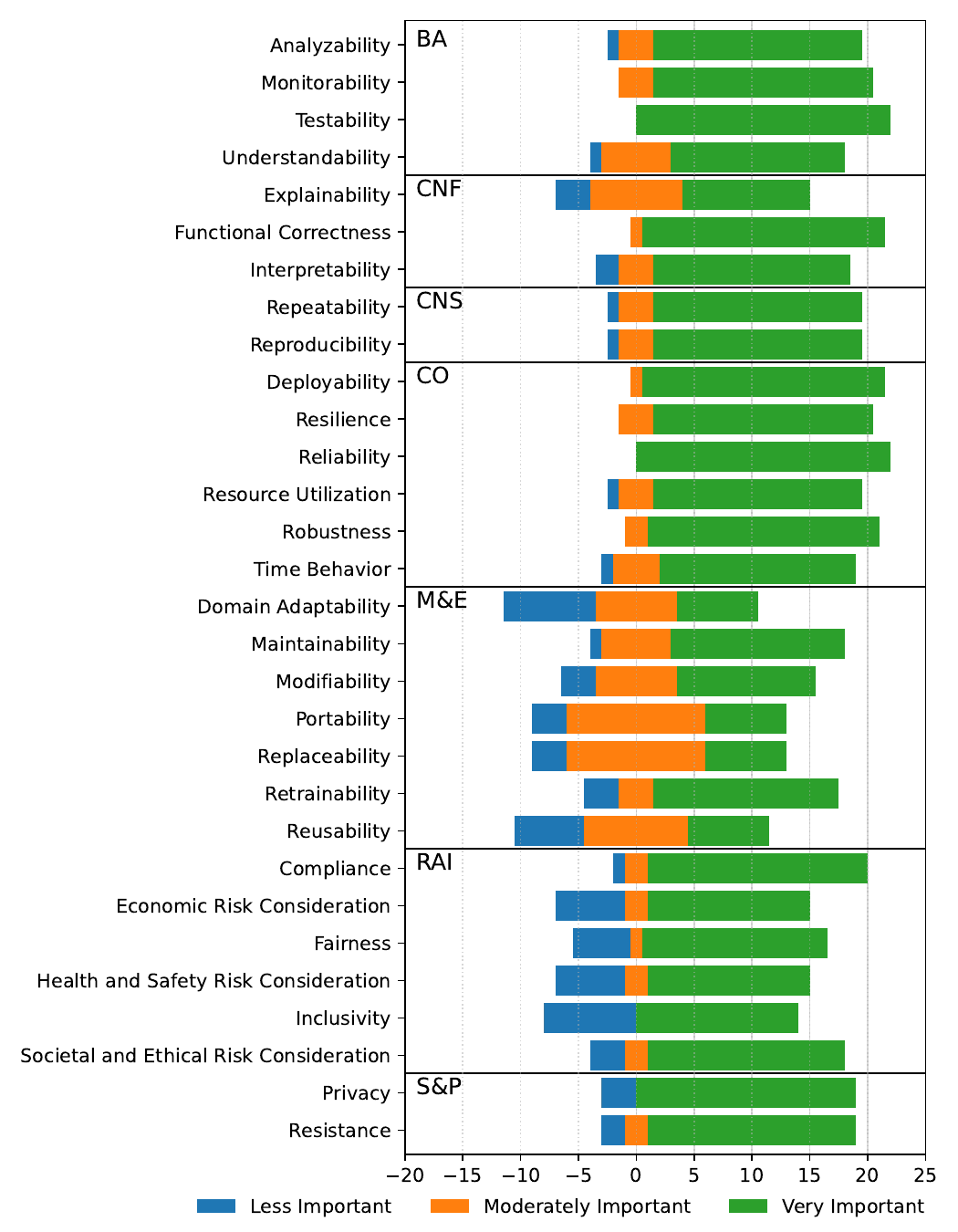}
    \caption{Importance of Testing QAs}
    \label{fig:survey-importance-counts}
  \end{subfigure}
  \hfill
  \begin{subfigure}[t]{0.45\textwidth}
    \centering
    \includegraphics[width=\textwidth]{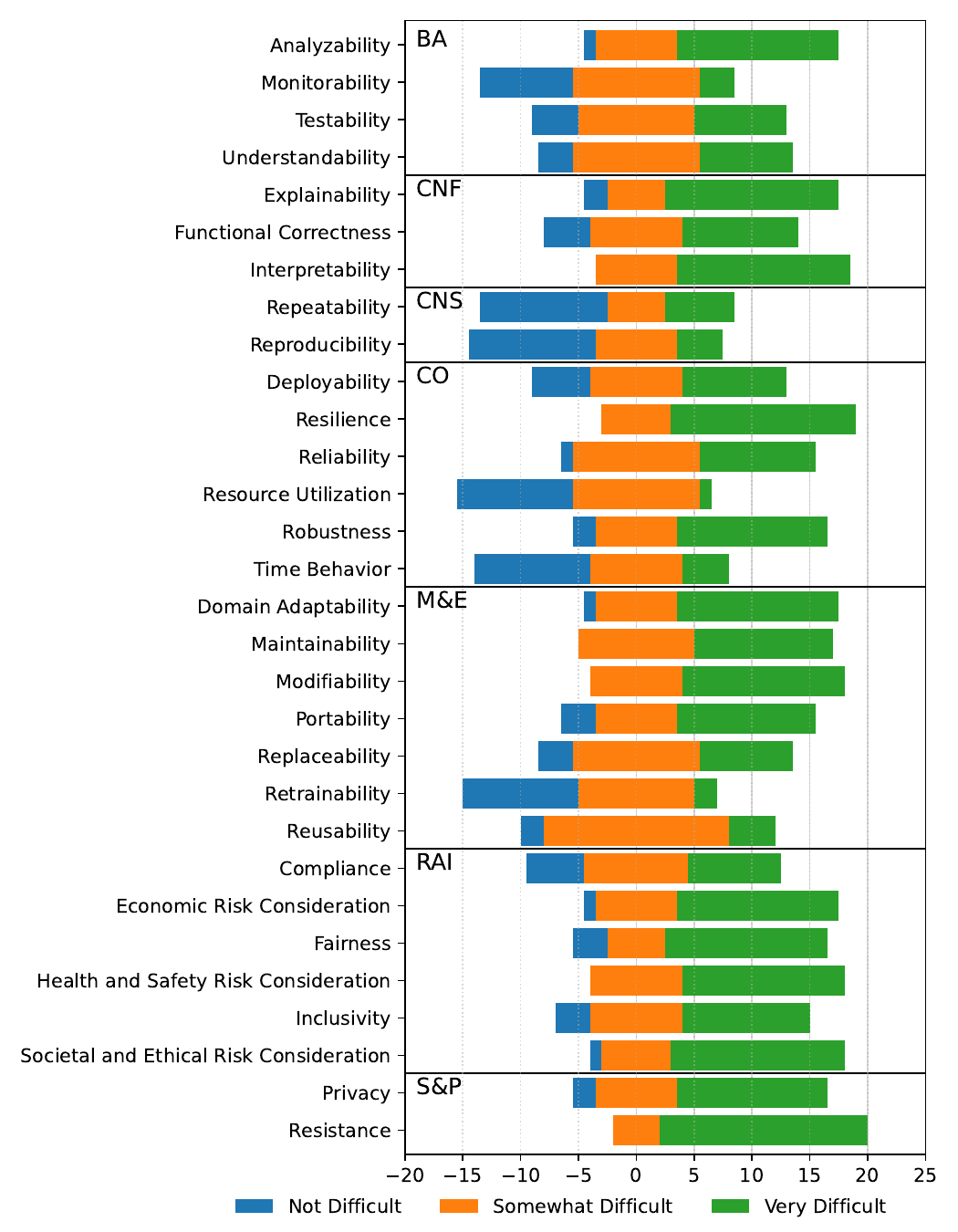}
    \caption{Difficulty of Testing QAs}
    \label{fig:survey-difficulty-counts}

  \end{subfigure} 
  \caption{Importance and Difficulty of Testing QAs in the Proposed Quality Model (BA=Behavior Analysis, CNF=Confidence, CNS=Consistency, CO=Continued Operation, M\&E=Maintenance and Evolution, RAI=Responsible AI, S\&P=Security and Privacy)}
  \label{fig:raw_results}
\end{figure*}
%


\subsection{Importance and Difficulty of Testing QAs (RQ3)}
\label{sec:survey-rq3)}
To assess RQ3, we asked participants to rate the importance and difficulty of each of the attributes in the proposed quality model using two five-point Likert scales: one for importance ranging from not important to very important, and one for difficulty ranging from very difficult to very easy. The results are shown in Figure \ref{fig:raw_results}.

From the importance perspective, reliability, testability, deployability, functional correctness, and robustness were rated as the top five of all QAs in the model. The bottom five included several of the QAs from the maintenance and evolution category: domain adaptability, portability, replaceability, reusability, and explainability.  Interestingly, from a difficulty perspective, all these lower importance QAs were ranked relatively higher on the difficulty scale by the participants. 

From the difficulty perspective, resistance was perceived by the participants as the most difficult to test, followed by resilience, societal and ethical risk considerations, interpretability, and explainability. The five QAs considered less difficult to test (\ie easier to test) were resource utilization, retrainability, monitorability, reproducibility, and time behavior. 

At the end of the survey we asked participants whether they agreed with the statement "Consideration of quality attributes in test and evaluation of ML components leads to early detection of a larger and more diverse set of problems that would otherwise not be detected until production." The participants unanimously strongly agreed (15/22) or agreed (7/22) with this statement. This perception is also somewhat reflected in the importance ratings. As can be seen in Figure ~\ref{fig:raw_results}(a), many of the QAs received an important or very important rating. The QAs with mixed ratings include those that fall under the \textit{Responsible AI} category, such as inclusivity, health and safety considerations, fairness, and economic risk considerations.


\section{Discussion}\label{sec:discussion}
In this section, we discuss the QAs that may be missing from our quality model based on our survey responses, the challenges of testing ML systems, and the implications of our results. 

\textbf{\textit{What is Missing from the Proposed Quality Model?}} One of the final questions in the survey was "Looking at the full quality model, what important quality attributes for ML components are missing?" Of the 15 participants who provided a response to this question, several expressed their agreement with the model and reflected that they do not see omissions based on their experience. Others provided feedback on the mapping of attributes to categories, which is reasonable. And others provided specific QAs that they believed were missing. Some of these corresponded to attributes that were covered under our definition of the QA, some corresponded to system-level attributes that we had already discarded, and others were relevant concrete examples of QAs that would be good to include in the QA definition as examples to further provide guidance to model developers, such as \textit{Energy Efficiency} which falls under our definition of \textit{Resource Utilization}. 


One QA to highlight is \textit{Scalability}, which was was mentioned in the survey in the top ten, both in the most important quality attributes to test according to Figure \ref{fig:RQ1-additionalQAs}, and in the answers to the missing QA question. In our analysis, we initially included scalability because it was included in the starting point ~\cite{ISO25010-2023,Habibullah2023,Chouliaras2023}. However, during the development of the quality model we unanimously viewed it primarily as a system property and removed it in the first round of Filtering (see Section \ref{fig:card-sorting}). Moreover, scalability was also not included in the AI-specific quality attributes in ISO 25059 ~\cite{ISO25059-2023}. However, it is interesting because testing the scalability of a model is more a test of the model-serving infrastructure and overall architecture of the system, and the model properties that are exercised for a model are a combination of time behavior (easier to scale if it is faster) and deployability (can scale by distributing your deployment). This is another good example of system-derived requirements.

\textbf{\textit{What Challenges Remain in Testing of ML Components?}} Another one of the final questions in the survey was "In general, what challenges do you foresee in test and evaluation of ML components?" The purpose of asking this question was to understand what practitioners still see as challenges and to potentially inform future work.

The most common challenges the participants mentioned were related to \textit{data-dependent behavior} and \textit{difficulty of establishing ground truth}, mainly in relation to LLMs and mismatches with operational environments once ML components are deployed. These are both well-known challenges in this field.

One participant summarized the data challenge from the perspective of both LLMs and traditional ML components: \textit{Access to ground truth information, particularly for LLM-based components is always a challenge. Frequently, the best information for this can be collected at prediction time or at execution time and is based on some action taken (or not) by the user, but this data is not always collected. Accounting for evolution of the problem an ML component is trying to solve is also missing from data collection and supporting test and evaluation.} Another participant focused on representativeness of test data compared to software testing: \textit{Test data quality is a big factor, as performance is tightly linked to data quality, distribution, and drift. For testing traditional ML models, data leakage can be stopped, but for testing LLMs, it is never guaranteed since we do not know what training data was used to train the LLM. Unlike software tests, where the result is pass or fail, testing ML models often involves looking at the distribution of results. This is where the "representativeness" of test data comes into play. Stress testing under scale, adversarial inputs, or noisy data is challenging.}

The participants also mentioned how uncertainties of the operational environment create challenges. For example, one participant summarized a challenge as: \textit{For our particular use case, the operational environment is very, very fluid and disjoint --- testing, evaluating, and generally understanding the nuances of the disjoint sets of environments the models need to operate in are a challenge.} Others reflected on the challenges of testing ML components due to growing demands, expressed as \textit{The demands placed on ML systems are growing and they are being pushed into wider domains and environments. The risks faced by ML systems are unknown unknowns in that there is a broad, potentially uncountable set of failures that are probabilistic in how they affect a system.}

\textbf{\textit{Implications of Results:}} A number of observations can be made based on our findings: 
\begin{itemize}
    \item The low importance rating of QAs related to \textit{Maintenance and Evolution} is not surprising, as most development efforts are focused on addressing capability needs today. However, this perspective may result in ML components, and consequently ML-enabled systems, that are difficult to maintain and evolve. As more organizations and teams move to maintenance and evolution of ML components currently under development, we envision that the importance of this category will increase.
    \item QAs related to \textit{Responsible AI} had the most mixed ratings, with the highest number of participants rating these QAs as not important to test. Several of these attributes were also rated as difficult to test. One potential explanation is that the return on investment of the testing effort is not high given the difficulty ratings. Another possible explanation may be that these attributes are assumed to be handled elsewhere \eg during data collection and preparation. In addition, despite the large amount of Responsible AI mandates and efforts across the globe (\eg US, EU, Australia), the reality is that practices for implementing and testing Responsible AI properties are less common \cite{Sadek2025}\cite{Schiff2020}. More research is needed to translate mandates into practice. 
    \item In general, the difficulty ranking for all QAs clustered in \textit{moderately difficult} to \textit{difficult} ratings, with fewer \textit{very difficult} responses, as can be seen in Figure ~\ref{fig:survey-difficulty-counts}. If the perception is that these QAs are not very difficult to test and that all participants agree that performing these tests results in earlier and improved detection of errors and issues, there is a huge opportunity for developing testing methods that can be easily integrated into ML component development and MLOps pipelines.  
    \item The \textit{Confidence} category shows a dichotomy. Although functional correctness is among the top three for importance to test, explainability and interpretability were lower in importance. However, these two attributes were also among the top ten as rated difficult to test. This finding may suggest that research has not yet closed the gap in improving explainability and interpretability to result in effective testing approaches. 
\end{itemize}

The main thing to consider is that ML components are not stand alone software, they work integrated into the bigger context of the software system that integrates them. As one respondent expressed: \textit{In addition to all of the AI specific correctness/robustness issues, you have all of the issues of testing a complex real-time system. This requires a lot of smart architecture decisions up front to enable the unit testing of the many nested ML components.} Our proposed QA model for ML components precisely fills this gap.

\section{Threats to Validity}\label{sec:threats}

\textit{External validity.} A potential threat to external validity is whether the proposed quality model can be generalized to different model types, such as LLMs. To mitigate this threat, we asked survey participants to indicate whether each QA applied to traditional models, LLMs or both. The results show that 70\% of the attributes apply to both traditional ML models and LLMs demonstrating broader applicability of our model. Another potential threat is whether the quality model generalizes to all types of ML-enabled systems. To mitigate this threat, we included the widely used ISO standard 25010 for software product quality in our starting point and complemented with work that specifically built on this standard for application to ML-enabled systems (see Section \ref{subsec:starting-point}). Finally, although the results are positive, the small sample size of survey participants also poses an external validity threat for generalizability of our model.

\textit{Internal validity}. A potential threat to internal validity is whether the card sorting process that we used to develop the quality model was rigorous enough to avoid personal bias. To mitigate this threat, we implemented a multi-step multi-iteration process that involved continuous discussion and conflict resolution, and added an additional team member in the final activity to bring in a new perspective and validation (see Section \ref{subsec:process}). Another potential threat is whether the literature search to identify complementary work to the standards was rigorous enough to identify all relevant work. As the goal was to identify examples instead of being exhaustive, and we were already familiar with the work in this area from previous work, to mitigate this risk, we defined very concrete search criteria looking specifically for studies proposing quality models for ML systems (see Section \ref{subsec:starting-point}). 
\section{Related Work}\label{sec:related-work}

Existing related work is mostly in requirements engineering and software architecture, identifying important quality attributes for AI-and ML-enabled systems or quality models for ML-enabled systems. 

Gezici \etal present the results of a systematic literature review (SLR) investigating the state of the art in software quality for AI-enabled systems, including the identification of quality attributes and quality models studied in the literature \cite{Gezici2022}. K{\"a}stner presents recommendations for QAs that should be specified and tested to move models from development to production \cite{Kastner2025}. Similarly, Pons \etal offer a set of QAs that are important for AI-enabled systems that operate in the public sector \cite{Pons2019} and Poth \etal present a list of relevant QAs based on an internal study at Volkswagen AG \cite{Poth2020}. Siebert \etal take more of a systems approach and present a list of QAs for trained model, data, environment, system, and infrastructure \cite{Siebert2020}\cite{Siebert2022}. Kuwajima \etal present an early attempt to adapt ISO 25010 to AI-enabled systems, which is the goal of ISO 25059, by extending the list of QAs to include ethical considerations \cite{Kuwajima2019}, similar to the work of Li \etal to define QAs for trustworthy AI-enabled systems \cite{Li2023}. These are all examples of work targeted at defining QAs for ML components or ML-enabled systems, but do not differentiate between QAs that are relevant for ML components and QAs that are relevant for ML-enabled systems. 

The work that is most closely related to ours is the two studies that served as a starting point for the development of the quality model for ML components \cite{Chouliaras2023}\cite{Habibullah2023}.  More recently, after developing our proposed quality model, Indykov \etal published the results of an SLR to identify common quality attributes for ML-enabled systems, architectural tactics to achieve those QAs, and potential trade-offs between quality attributes \cite{Indykov2025}. The quality model produced by this study overlaps with all the proposed quality models, including ours, and after careful review, it would not have changed our proposed model based on our methodology and discussions. Something to note that is different in this quality model is the inclusion of data quality as a QA. Although we agree that this is a very important aspect of quality, as also confirmed by our survey respondents as a significant challenge, we consider that data-related aspects require its own separate data quality model and is the next step of this work (see Section \ref{sec:summary}). Regardless, all quality models proposed by these studies also have the problem of mixing QAs that apply at the system level with those that apply at the ML component level, especially when looking at QAs related to usability and security. 

To the best of our knowledge, this is the first empirically developed and validated quality model that focuses on testable quality attributes for ML components and, therefore, enables their proper specification and testing in the context of an ML-enabled system.  
\section{Summary and Next Steps}\label{sec:summary}

We presented a quality model for ML components that guides their development and testing, especially in situations in which (1) model development occurs in separate teams or organizations from those developing the ML-enabled system, or (2) model development is performed by data scientists not familiar with quality attributes. The model has been successfully integrated into MLTE, an open source tool for test and evaluation of ML components as a resource to guide elicitation and negotiation of system-derived requirements. 

Next steps for this work include (1) larger-scale evaluation of the quality model, (2) user study of the quality model as part of the MLTE tool, and (3) development of a data quality model using the same methodology, with the goal of integration into MLTE.




\begin{acks}
This material is based upon work supported by the Department of War under Air Force Contract No. FA8702-15-D-0002 with Carnegie Mellon University for the operation of the Software Engineering Institute, a federally funded research and development center (DM26-0066).
\end{acks}

\bibliographystyle{ACM-Reference-Format}
\bibliography{references}

\end{document}